# Fundamental Antisymmetric Mode Acoustic Resonator in Periodically Poled Piezoelectric Film Lithium Niobate

Omar Barrera[1], Jack Kramer[1], Ryan Tetro[2], Sinwoo Cho[1], Vakhtang Chulukhadze[1], Luca Colombo[2], and Ruochen Lu[1]
[1]The University of Texas at Austin, TX, USA
[2]Northeastern University, MA, USA

*Abstract*— Radio frequency (RF) acoustic resonators have long been used for signal processing and sensing. Devices that integrate acoustic resonators benefit from their slow phase velocity ($v_p$), in the order of 3 to 10 km/s, which allows miniaturization of the device. Regarding the subject of small form factor, acoustic resonators that operate at the so-called fundamental antisymmetric mode (A0), feature even slower $v_p$ (1 to 3 km/s), which allows for smaller devices. This work reports the design and fabrication of A0 mode resonators leveraging the advantages of periodically poled piezoelectricity (P3F) lithium niobate, which includes a pair of piezoelectric layers with opposite polarizations to mitigate the charge cancellation arising from opposite stress of A0 in the top and bottom piezoelectric layers. The fabricated device shows a quality factor ($Q$) of 800 and an electromechanical coupling ($k^2$) of 3.29, resulting in a high figure of merit (FoM, $Q \cdot k^2$) of 26.3 at the resonant frequency of 294 MHz, demonstrating the first efficient A0 device in P3F platforms. The proposed A0 platform could enable miniature signal processing, sensing, and ultrasound transducer applications upon optimization.

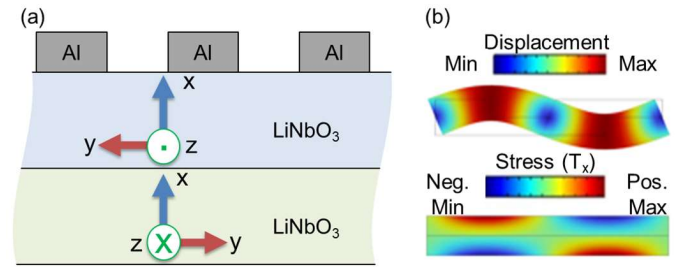

Fig. 1. (a) LiNbO$_3$ dual layer resonator stack (b) Simulated A0 displacement and stress profiles.

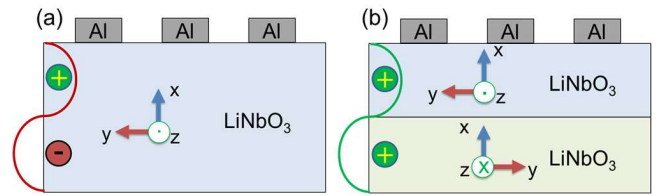

Fig.2. (a) Stress antinode formation in single layer A0 mode leading to charge cancellation (b) Fully harnessing piezoelectric convertion using P3F dual layer stack.

## I. INTRODUCTION

There is a never-ending demand for smaller devices in the radio-frequency (RF) commercial electronics industry. Not only are small form factor devices required, but the performance of the components is expected to remain comparable. To this end, acoustic wave resonators have achieved great success in applications such as miniature sensors, high quality factor ($Q$) resonators and front-end filters for mobile applications [1], [2]. Acoustic resonators utilize the principle of piezoelectricity to transduce energy back and forth between electrical signals and mechanical oscillations. Therefore, it is possible to perform the sensing or signal processing in the mechanical domain, which has the advantage of slow phase velocities ($v_p$) in the order of 3 to 10 km/s, several orders of magnitude lower than the electrical counterparts [3], [4]. Among different types of acoustic vibrations, resonators working at the fundamental symmetric mode (S0) [5]–[8], the fundamental shear horizontal mode (SH0) [9]–[13], first-order antisymmetric (A1) [14]–[19], and first-order symmetric (S1) [20]–[27] have been widely studied. Such devices have demonstrated strong potential and are still heavily investigated. However, it is not easy to further tune down the $v_p$ of these modes below 3 km/s. This factor limits the range of device size reduction.

Another mode, namely the fundamental antisymmetric mode (A0), features even slower $v_p$ of less than 2 km/s [28]. Thus, exciting this mode opens the possibility of further device dimensions downsizing. However, an efficient excitation of the A0 mode with high figures of merit is not trivial. In the case of a single film, due to the nature of the motion, stress antinodes with opposite signs develop at the top and bottom of the piezoelectric layer. This effect leads to charge cancellation, adversely affecting the achievable electromechanical coupling ($k^2$), thus lowering the figure of merit (FoM, $Q \cdot k^2$). Adding a passivation silicon dioxed (SiO$_2$) layer can partially mitigate this issue, but this process itself also brings about degraded $k^2$ and $Q$ [29], [30]. Recently, resonators leveraging the periodically poled piezoelectric (P3F) effect have been proposed to exploit over-moded lithium-niobate (LiNbO$_3$) resonators [31]–[33]. Over-moded resonators also suffer from charge cancellation effects, and P3F films have the potential to help mitigate the issues. However, studies on enhancing the performance of A0 mode device with P3F LiNbO₃ are not reported.

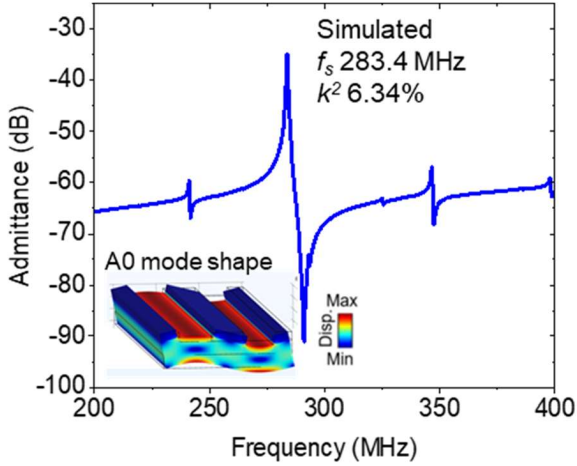

Fig. 3. Simulated admittance of the unit cell resonator and A0 flexural displacement mode shape.

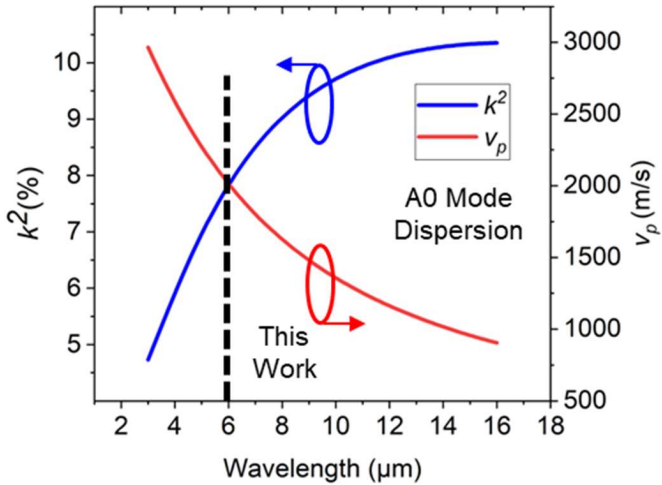

Fig. 4. Simulated $k^2$ and $v_p$ dispersion curves against changes in the wavelength of a unit cell A0 resonator.

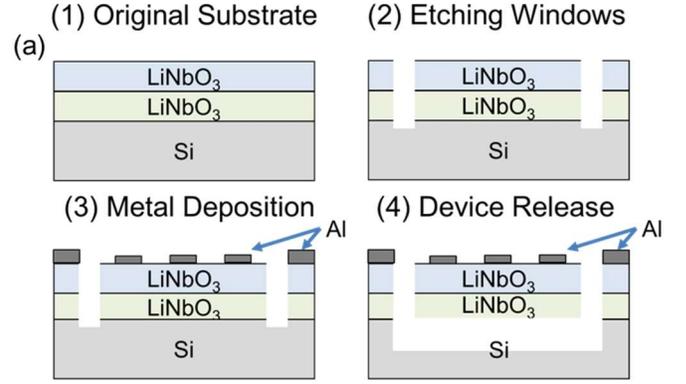

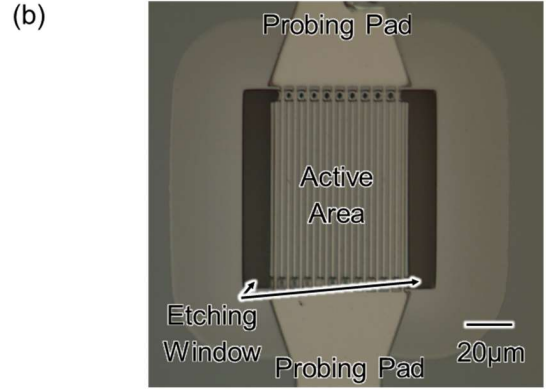

Fig.5. (a) Step-by-step fabrication process diagram of the A0 resonators (b) Optical image of the full fabricated device.

In this work, we implemented the first A0 mode resonators in a P3F thin-film. The devices are built on a dual-layer X-cut LiNbO₃ film, achieving a $Q$ of 800, $k^2$ of 3.29% and a high FoM of 23.6 at 294 MHz, while maintaining a $v_p$ of 1800 m/s, 2 to 5 times smaller than S0 modes. The proposed A0 platform could enable miniature signal processing, sensing, and ultrasound transducer applications upon optimization.

## II. DESIGN AND SIMULATION

The device structure is formed by 2 layers of X-cut LiNbO₃, each layer with a thickness of 550 nm and the y-orientation in opposite direction [Fig. 1(a)]. The displacement and stress profiles for a canonical A0 shape motion are shown in Fig. 1(b). The displacement is purely vertical and it has the characteristic flexural bending associated with the fundamental mode. The lateral stress antinodes, as expected, are at the top and bottom of the thin film and have opposite signs. In conventional single-layer piezoelectric materials, it is hard to harness the generated charge with a single pair of top electrodes, as the sign of charges in the upper and lower sections will perfectly cancel [Fig. 2(a)]. However, in P3F LiNbO₃, due to the different signs in the piezoelectric coefficient ($e_{11}$), we will fully harness the piezoelectricity [Fig. 2(b)].

The resonator is excited by aluminum (Al) interdigitated electrodes (IDT) sitting on top of the stack, the wavelength ($\lambda$) is chosen as 6 μm, and a constant duty cycle of 50% is maintained. The metal thickness is selected as 350 nm. A single unit cell of the proposed design was simulated using COMSOL finite element analysis (FEA) to verify the mode shape and the effect of the P3F film stack, the results are shown in Fig. 3. The device exhibits an electromechanical coupling ($k^2$) of 6.34% at the resonant frequency of 283.4 MHz.

A comparison between $k^2$ and $v_p$ against changes in the $\lambda$ is plotted in Fig. 4. It can be observed that at 6 μm the $v_p$ is 1,800 ms/s, thus confirming the potential of P3F resonators to realize miniaturized devices. Additionally, the intersection between $k^2$ and $v_p$ also validates the selection of $\lambda$ as it offers a good tradeoff between these 2 parameters.

## III. FABRICATION AND MEASUREMENT

The fabrication process starts with a dual layer X-cut LiNbO₃, total thickess of 1.1 μm, thin film transferred on top of a silicon (Si) carrier wafer. The material stack is provided by NGK

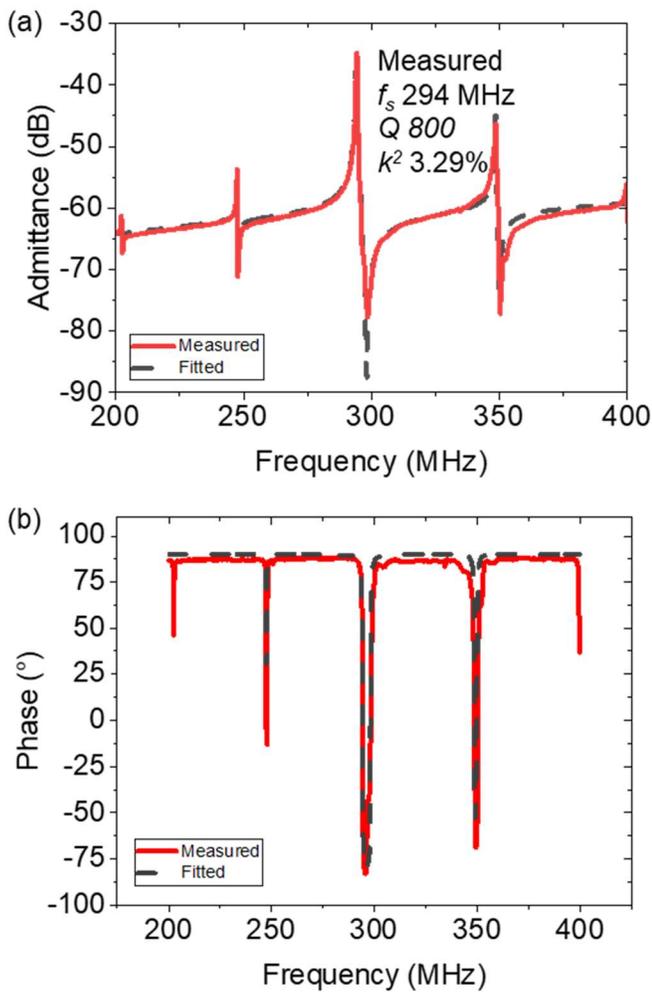

Fig.6. (a) Measured and fitted response of the A0 resonator in (a) admittance and (b) phase.

Insulators, Inc. First, opening windows for device release are patterned on top using traditional lithography techniques. Next, ion milling is used to etch through the LiNbO$_3$ stack deep into the silicon wafer. Afterwards, the features for metal interconnects are patterned using electron beam lithography, and e-beam evaporation is used to deposit 350 nm of Al for the buslines and electrodes. Finally, the resonators are released using xeon difluoride (XeF$_2$) for silicon etching. The step-by-step fabrication process is depicted in Fig. 5(a). An optical image of the final suspended device is shown in Fig. 5(b), the large lateral etch windows perform most of the isotropic release process, while the small etch windows between electrodes help confine acoustic energy in the active area.

The fabricated device is measured using a Keysight vector network analyzer (VNA), and the measured data is then fitted using a modified Butterworth-Van Dyke model. The main resonance is located at 294 MHz, in good agreement with the expected behaviour from the simulation. The device exhibits a $Q$ of 800 and a $k^2$ of 3.29%, resulting in a high FoM of 26.3. The reduced $k^2$ is due to crystal orientation misalignment of the top and bottom layers during the bonding, which could be improved in future works. $Q$ could be potentially enhanced by minimizing lattice damage during thin film transfer. A possible approach to maintaining good lattice properties could involve using an intermediate amorphous silicon (a-Si) layer between the LiNbO$_3$ stack and carrier susbtrate [34]. The measurements, fitting and extracted main parameters are plotted in Fig. 6 (a) and (b), respectively.

IV. CONCLUSION

In this work, we report the first A0 mode resonator leveraging a P3F X-cut LiNbO$_3$ bi-layer stack. The device shows a Q of 800 and $k^2$ at the resonant frequency of 294 MHz. The resulting FoM of 23.6 with a slow $v_p$ of 1,800 m/s could enable miniature scale resonators for signal processing, sensing, and ultrasound transducer applications.

ACKNOWLEDGMENT

The authors would like to thank the funding support from the DARPA COFFEE program and Dr. Ben Griffith for the helpful discussion.